\documentclass{ifacconf}

\usepackage{graphicx}
\usepackage{natbib}
\usepackage{amsmath}
\usepackage{amssymb}
\usepackage{tikz}
\usetikzlibrary{shapes.geometric, arrows.meta, positioning, fit, backgrounds, calc}

\usepackage{xcolor}
\usepackage{colortbl}
\usepackage{booktabs}
 
\definecolor{LowColor}{HTML}{F09595}
\definecolor{MedColor}{HTML}{FAC775}
\definecolor{HighColor}{HTML}{97C459}
\definecolor{EarlyColor}{HTML}{F09595}
\definecolor{PilotColor}{HTML}{FAC775}
\definecolor{BenchColor}{HTML}{5DCAA5}

\let\oldthebibliography\thebibliography
\renewcommand{\thebibliography}[1]{%
  \oldthebibliography{#1}%
  \setlength{\itemsep}{0pt}%
  \setlength{\parskip}{0pt}}

\usepackage{titlesec}
\titleformat{\section}[block]{\normalfont\normalsize\centering}{\thesection.}{0.5em}{\MakeUppercase}
\titleformat{\subsection}[block]{\normalfont\normalsize\itshape}{\thesubsection}{0.5em}{}
\titlespacing{\section}{0pt}{1pt plus 1pt minus 1pt}{0pt plus 1pt minus 1pt}
\titlespacing{\subsection}{0pt}{0pt plus 1pt minus 1pt}{-1pt plus 1pt minus 1pt}

\usepackage{caption}
\begin{document}

\setlength{\textfloatsep}{8pt plus 2pt minus 2pt}
\setlength{\floatsep}{5pt plus 1pt minus 1pt}
\setlength{\intextsep}{6pt plus 2pt minus 2pt}

\begin{frontmatter}

\title{Large Language Models in Process Systems Engineering:\\ Opportunities, Architectures, and Industrial Deployment Challenges\thanksref{footnoteinfo}}

\thanks[footnoteinfo]{This survey accompanies the IFAC World Congress 2026 Workshop on ``Applications of Large Language Models in Process Control.''}

\author[UBC]{R. Bhushan Gopaluni}
\author[Syris]{Vidya Kotamraju}
\author[StGeorge]{Syon Bhushan}

\address[UBC]{Department of Chemical and Biological Engineering, University of British Columbia, Vancouver, BC, Canada (e-mail: bhushan.gopaluni@ubc.ca)}
\address[Syris]{Syris AI Systems; Adjunct Faculty, University of British Columbia, Vancouver, BC, Canada (e-mail: vidya@syris.ai)}
\address[StGeorge]{St. George's High School, Vancouver, BC, Canada (e-mail: syonisms@gmail.com)}

\begin{abstract}
Large Language Models (LLMs) have rapidly emerged as tools of interest across engineering disciplines, and Process Systems Engineering (PSE) is no exception. This survey provides a systematic review of LLM applications in PSE, organizing the literature into seven categories: (1) process design and engineering, (2) molecular design and synthesis, (3) process modeling and simulation, (4) time-series forecasting, (5) optimization and scheduling, (6) process control, and (7) fault detection and diagnosis. For each category, we summarize the state of the art, identify common methodological approaches, and critically assess demonstrated capabilities versus aspirational claims. We find that LLMs show genuine promise for tasks involving natural language, including querying documentation, synthesizing unstructured knowledge, and enabling flexible human-machine interaction. However, applications requiring real-time execution, constraint satisfaction, or formal safety guarantees remain challenging. We conclude by identifying open problems and productive research directions for the PSE community.
\end{abstract}


\end{frontmatter}

\section{Introduction}

Large Language Models (LLMs) have attracted substantial attention across virtually every technical domain since the public release of ChatGPT in November 2022. Process Systems Engineering (PSE), encompassing the design, modeling, control, optimization, and monitoring of chemical and process systems, has seen a growing body of work exploring LLM applications. This survey provides a systematic review of this emerging literature, with particular attention to what new human--system interaction and orchestration capabilities LLMs enable for process operations. We argue that they are best positioned as supervisory layers rather than replacements for classical control.

The process industries face a persistent challenge: the gap between data abundance and actionable insight. Modern plants can generate large volumes of sensor and event data from high-frequency historians, yet operators often struggle to synthesize this information with maintenance histories, operating procedures, and engineering knowledge scattered across documents, databases, and institutional memory. Traditional approaches such as rule-based expert systems, statistical process monitoring, and Model Predictive Control (MPC) excel at their designated functions but operate in silos. An MPC controller optimizes setpoints but cannot explain its reasoning to an operator; a fault detection system flags anomalies but cannot connect them to similar historical events described in maintenance logs; a process historian stores years of data but requires specialized queries to extract meaning. Distributed Control Systems (DCS), Supervisory Control and Data Acquisition (SCADA) systems, and Manufacturing Execution Systems (MES) each serve important functions but rarely share information seamlessly. LLMs offer, for the first time, a plausible path toward systems that bridge these silos through natural language.

The appeal of LLMs for process operations stems from several distinctive capabilities. First, LLMs process and generate natural language with unprecedented fluency, enabling interfaces where operators interact with plant systems conversationally rather than through rigid command structures. An operator asking ``Why did the reactor temperature spike during last night's grade transition?'' could receive an answer synthesizing real-time data, historical trends, and documented procedures. This type of query is difficult to answer with traditional systems. Second, LLMs exhibit in-context learning: the ability to sometimes adapt to new tasks through examples provided in the prompt, without retraining. This flexibility can enable rapid prototyping of domain-specific applications, though reliability varies by task and prompt design. Third, modern LLMs can invoke external tools such as simulators, databases, and optimization solvers based on natural language instructions, enabling agentic workflows where the model reasons about which tools to apply and how to interpret their outputs.

Beyond interfacing, LLMs bring qualitative reasoning capabilities that complement quantitative methods. Process control has traditionally focused on what can be measured and modeled mathematically: temperatures, pressures, flows, compositions. Yet plant operations involve substantial qualitative knowledge: heuristics for startup sequences, experience-based rules for troubleshooting, operator intuition about ``how the plant feels.'' This knowledge resists formalization but is captured, imperfectly, in operating procedures, shift logs, and incident reports. This is precisely the textual data that LLMs can process. The prospect of systems that combine rigorous model-based control with flexible reasoning over operational knowledge is genuinely novel.

Our objectives in this survey are threefold. First, we organize the scattered literature into coherent categories that reflect the structure of PSE research: process design, molecular synthesis, modeling and simulation, time-series forecasting, optimization and scheduling, process control, and fault detection and diagnosis. Second, for each category, we summarize what has been demonstrated, what methods are employed, and what claims require further validation. Third, we offer a critical assessment of where LLMs genuinely advance PSE capabilities versus where enthusiasm may have outpaced evidence. We include peer-reviewed articles and relevant preprints through early 2026; given the rapid pace of development in this field, some cited works may not yet have undergone formal peer review.

\textit{Scope clarification}: We survey LLM applications across the entire PSE lifecycle, from molecular design through plant operations. Chemistry and materials sections are included as upstream enablers, since molecular property prediction and retrosynthesis directly inform process design decisions. The emphasis throughout is on applications relevant to industrial process systems rather than pure chemistry or materials science.

\textit{Survey method}: This is an expert narrative survey rather than a formal systematic review. Literature was identified through the authors' ongoing research activities, citation tracking, and targeted searches of recent publications. We prioritize representative works that illustrate key themes over comprehensive enumeration. Preprints are included where they represent significant developments not yet available in peer-reviewed form.

We find that the most compelling applications leverage LLMs' core strengths: processing natural language, synthesizing information from diverse textual sources, and enabling flexible interaction between humans and technical systems. Applications that treat LLMs as replacements for purpose-built numerical methods (optimizers, controllers, simulators) are generally less convincing. A model predictive controller, once configured, executes in milliseconds with explicit constraint handling; an LLM queried for the same control action requires seconds, provides no comparable guarantees, and may hallucinate. The distinction matters for directing future research productively: LLMs as orchestrators and interfaces rather than as substitutes for validated numerical methods.

The survey is organized as follows. Section~\ref{sec:background} provides background on LLM architectures and capabilities, with particular attention to their potential as time-series modeling tools. Section~\ref{sec:design} reviews applications in process design and engineering. Section~\ref{sec:molecular} covers molecular design and synthesis. Section~\ref{sec:modeling} addresses process modeling and simulation. Section~\ref{sec:forecasting} examines the emerging area of LLM-based time series forecasting, including foundation models. Sections~\ref{sec:optimization}--\ref{sec:fdd} review optimization, control, and fault diagnosis applications respectively. Section~\ref{sec:challenges} discusses cross-cutting challenges including hallucination, latency, and validation, and Section~\ref{sec:future} outlines future directions for the PSE community.

\begin{figure*}[t]
\centering
\vspace{-6pt}
\begin{tikzpicture}[
    node distance=0.5cm,
    box/.style={rectangle, draw, rounded corners, minimum width=5.5cm, minimum height=0.7cm, align=center, font=\small},
    widebox/.style={rectangle, draw, rounded corners, minimum width=7cm, minimum height=0.7cm, align=center, font=\small},
    arrow/.style={-{Stealth[scale=0.8]}, thick}
]
\node[box, fill=blue!15] (human) {Operators / Engineers};

\node[widebox, fill=orange!20, below=of human] (llm) {LLM Reasoning \& Interface Layer};

\node[box, fill=green!15, below left=0.5cm and -0.5cm of llm] (knowledge) {Knowledge Systems};
\node[box, fill=green!15, below right=0.5cm and -0.5cm of llm] (simulation) {Simulation / Digital Twin};

\node[box, fill=yellow!15, below=0.5cm of knowledge] (optimization) {MPC / Optimization};
\node[box, fill=yellow!15, below=0.5cm of simulation] (analytics) {Fault Detection};

\node[widebox, fill=gray!20, below=0.9cm of $(optimization)!0.5!(analytics)$] (dcs) {DCS / SCADA / Physical Process};

\draw[arrow, <->] (human) -- node[right, font=\footnotesize] {NL} (llm);
\draw[arrow, <->] (llm) -- (knowledge);
\draw[arrow, <->] (llm) -- (simulation);
\draw[arrow, ->] (knowledge) -- (optimization);
\draw[arrow, ->] (simulation) -- (analytics);
\draw[arrow, <->] (optimization) -- (dcs);
\draw[arrow, <->] (analytics) -- (dcs);

\end{tikzpicture}
\caption{LLM-enabled plant intelligence architecture. LLMs serve as a reasoning and interface layer, orchestrating interactions between operators, knowledge systems, and validated numerical methods. Classical controllers (MPC, PID) retain real-time execution authority with formal guarantees; LLMs provide natural language interaction, knowledge synthesis, and high-level coordination. NL = Natural Language.}
\vspace{-10pt}
\label{fig:architecture}
\end{figure*}
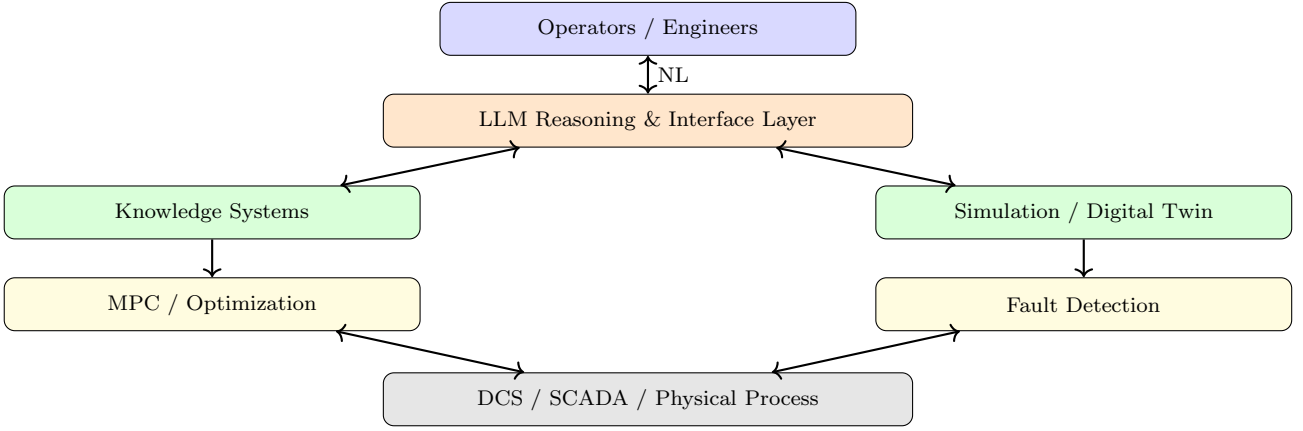

Figure~\ref{fig:architecture} illustrates the conceptual architecture underlying our analysis. The central thesis is that LLMs are best positioned as a cognitive layer above, not replacing, engineered control systems. Table~\ref{tab:comparison} summarizes where each approach excels.


\begin{table}[t]
\centering
\caption{Complementary strengths of LLMs and classical methods.}
\label{tab:comparison}
\footnotesize
\begin{tabular}{lcc}
\hline
\textbf{Task} & \textbf{LLMs} & \textbf{Classical Methods} \\
\hline
Natural language interaction & \checkmark & -- \\
Knowledge synthesis & \checkmark & -- \\
Real-time control ($<$100ms) & -- & \checkmark \\
Constraint guarantees & -- & \checkmark \\
Safety certification & -- & \checkmark \\
Numerical optimization & -- & \checkmark \\
Flexible reasoning & \checkmark & -- \\
Physics-based simulation & -- & \checkmark \\
\hline
\end{tabular}
\end{table}

Table~\ref{tab:taxonomy} provides a taxonomy of LLM applications in PSE, classifying by LLM role, coupling with physics-based methods, and validation level. We note that the literature identified in the paper is unevenly distributed across application domains partly due to space constraints and the authors' experience and knowledge. 

\begin{table}[t]
\centering
\caption{Taxonomy of LLM applications in PSE by role, engineering methods and validation.}
\label{tab:taxonomy}
\footnotesize
\begin{tabular}{@{}llll@{}}
\toprule
\textbf{Application} & \textbf{LLM Role} & \textbf{Engg. Methods} & \textbf{Validation} \\
\midrule
Flowsheets & Generation & Simulator & Benchmark \\
P\&IDs & Query / Gen & Knowledge graph & Lab demo \\
Retrosynthesis & Planning & Robotic platform & Lab demo \\
Forecasting & Time-series model & Hybrid ML & Benchmark \\
Optimization & Orchestration & Optim.\ solver & Simulation \\
Control & Supervisory reasoning & MPC / PID & Lab demo \\
Fault diag. & Diag.\ reasoning & Retrieval systems & Simulation \\
HAZOP & Generation & Domain KB & Expert rev. \\
\bottomrule
\end{tabular}
\vspace{4pt}
\parbox{\columnwidth}{\footnotesize Validation categories are descriptive rather than ordinal.}
\end{table}

\section{Background} \label{sec:background}

\subsection{Transformer Architecture}

Modern LLMs are built on the transformer architecture introduced by \citet{vaswani2017attention}. Unlike recurrent neural networks that process sequences token-by-token, transformers process entire sequences in parallel through self-attention mechanisms. This parallelism enables efficient training on massive datasets and captures long-range dependencies that RNNs struggle with due to vanishing gradients.

The core innovation is the attention mechanism, which computes relevance scores between all pairs of positions in a sequence. Given an input sequence, the model learns three projections for each token: queries ($Q$), keys ($K$), and values ($V$). Attention scores are computed as:
\begin{equation}
\text{Attention}(Q,K,V) = \text{softmax}\left(\frac{QK^T}{\sqrt{d_k}}\right)V
\end{equation}
where $d_k$ is the key dimension. This allows each output position to attend selectively to relevant input positions, regardless of distance. Multi-head attention extends this by learning multiple attention patterns in parallel, capturing different types of relationships.

Transformers stack these attention layers with feed-forward networks, layer normalization, and residual connections. GPT-style models use decoder-only architectures with causal (unidirectional) attention: each position can only attend to previous positions, enabling autoregressive generation where tokens are produced sequentially, each conditioned on all previous tokens.

\subsection{Scaling and Emergent Capabilities}

The remarkable capabilities of modern LLMs emerged from scaling: larger models trained on more data exhibit qualitatively different behaviors \citep{kaplan2020scaling}. GPT-3 (175 billion parameters) demonstrated few-shot learning, performing tasks from a handful of examples in the prompt, that smaller models could not match. Subsequent models (GPT-4, Claude) have shown continued capability gains, including complex reasoning, code generation, and multimodal understanding. The opportunities and risks of such foundation models have been extensively analyzed \citep{bommasani2021opportunities}.

Reinforcement Learning from Human Feedback (RLHF) further shapes model behavior \citep{ouyang2022training}. After initial pre-training on text prediction, models are fine-tuned using human preferences: evaluators rank model outputs, and the model learns to produce responses humans rate highly. This process improves instruction-following, reduces harmful outputs, and generally aligns model behavior with user intentions.

\subsection{Capabilities Relevant to PSE}

Several LLM capabilities are particularly relevant for PSE:

\textit{Natural language understanding}: LLMs parse complex queries, extract intent, and generate coherent responses, enabling interfaces to plant documentation and historical records.

\textit{In-context learning}: LLMs adapt to new tasks through examples in the prompt, without parameter updates, enabling rapid customization to plant-specific conventions.

\textit{Tool use}: Tool-augmented agent patterns such as ReAct \citep{yao2023react} enable models to decide when to call external functions including simulators, databases, and calculators. Chain-of-thought prompting \citep{wei2022chainofthought} and reflexion \citep{shinn2023reflexion} further enhance reasoning.

\textit{Retrieval-augmented generation}: RAG techniques \citep{lewis2020rag} ground LLM outputs in retrieved documents, reducing hallucination and enabling access to domain-specific knowledge.

\textit{Multimodal processing}: Recent models process images alongside text, enabling interpretation of P\&IDs, trend plots, and equipment photographs.

\textit{Code generation}: LLMs generate and debug code, enabling automated creation of data processing scripts and data analysis routines.

\subsection{Limitations}
Several fundamental limitations constrain LLM applications in PSE and must inform system design:

\textit{Hallucination}: LLMs generate plausible but incorrect content with no internal reliability signal. The model produces confident statements about reactor conditions, equipment specifications, or recommended actions that may be fabricated. Hallucination is a known failure mode of current LLMs, a consequence of next-token generation using probabilistic correlations. While mitigations such as RAG reduce frequency, safety-critical designs should assume hallucination can still occur and require external verification of all consequential outputs.

\textit{Latency}: State-of-the-art models require hundreds of milliseconds to seconds per response when accessed via cloud APIs. Local deployment of smaller models reduces latency but also reduces capability. For control loops requiring response times on the order of tens of milliseconds, current LLM inference is too slow; supervisory tasks operating on minutes-to-hours timescales are more plausible. This constraint may ease with hardware advances and model optimization, but it fundamentally shapes which applications are feasible today.

\textit{No formal guarantees}: Model predictive control provides constraint satisfaction (when feasible); Lyapunov analysis certifies stability (under modeling assumptions); formal verification proves safety properties. LLMs provide no comparable analysis frameworks. We cannot prove an LLM-based system will never recommend an unsafe action, will always detect a particular fault, or will respond consistently to identical inputs. For applications requiring such guarantees, LLMs cannot be the sole decision-maker.

\textit{Opacity}: Despite extensive interpretability research, we cannot fully explain why LLMs produce particular outputs. Attention weights provide partial insight but do not constitute a complete explanation. This opacity complicates debugging, validation, and regulatory approval in industries where explainability is required.

\textit{Training data dependence}: LLM capabilities reflect their training data. Models trained primarily on web text may lack deep knowledge of specialized process equipment, proprietary control strategies, or plant-specific conventions. Fine-tuning and retrieval-augmented generation can address gaps, but require careful curation of domain-specific data.

These limitations are not reasons to avoid LLMs but should inform application design. The appropriate role for LLMs is where their strengths (language, reasoning, knowledge synthesis, flexibility) provide value while traditional methods handle verification, real-time execution, and safety-critical decisions. The architectural pattern of LLMs as reasoning engines that propose actions validated by simulation before execution, or as interpreters that explain decisions made by conventional controllers, leverages respective strengths while mitigating respective weaknesses.

\section{Process Design and Engineering} \label{sec:design}

Process design translates requirements into Block Flow Diagrams (BFDs), Process Flow Diagrams (PFDs), Piping and Instrumentation Diagrams (P\&IDs), and control architectures. This labor-intensive workflow, spanning conceptual design through detailed engineering, has attracted substantial LLM research across multiple fronts.

\subsection{Flowsheet Representation and Language}

A foundational challenge in applying machine learning to process design is representing flowsheets in formats that preserve process connectivity while remaining amenable to computational models. The Simplified Flowsheet Input Line Entry System (SFILES) provides text-based flowsheet representation similar to SMILES for molecules \citep{vogel2023sfiles}. \citet{mann2024esfiles} extended this to eSFILES, a multi-level hierarchical representation combining text strings, hypergraphs, and ontologies to capture connectivity, process groups, and operational parameters at varying abstraction levels.

These representations enable transformer-based approaches. \citet{vogel2023flowsheetautocompletion} demonstrated flowsheet autocompletion using generative language models: pre-training on synthetic flowsheets to learn SFILES grammar, then fine-tuning on real topologies. The approach provides engineers recommendations during interactive synthesis, analogous to code completion in software development. \citet{balhorn2024autocorrection} explored LLM-based autocorrection of chemical process flowsheets, detecting and proposing fixes for common design errors. \citet{theisen2023digitization} developed deep CNNs to recognize flowsheets within literature images, enabling automated mining of process knowledge from publications and patents.

\subsection{Control Structure Prediction}

\citet{hirtreiter2024toward} framed control structure prediction as a translation problem: a process flow diagram without controllers is mapped to a corresponding diagram with control structures. Using transformer models, the authors pre-trained on 100,000 synthetic diagrams and then applied transfer learning to industrial examples, reporting 74.8–89.2\% top-5 accuracy. This work illustrates how text-based flowsheet representations such as SFILES can support machine-learning approaches to control-structure synthesis. Graph-based variants, in which graph neural networks encode the process topology before generating SFILES-like outputs, provide an alternative that is less sensitive to the ordering of units in the input sequence.

\subsection{P\&ID Interaction and Natural Language Interfaces}

\citet{alimin2025talking} developed natural language interfaces for querying P\&IDs. P\&IDs are converted to knowledge graphs using the DEXPI data exchange standard via the pyDEXPI Python framework \citep{goldstein2025pydexpi}, then integrated with LLMs through graph-based retrieval-augmented generation. Engineers can query specifications (``What is the design pressure of V-101?'') with responses grounded in structured diagram data, significantly mitigating hallucination compared to direct LLM queries.

\citet{gowaikar2024agentic} explored agentic P\&ID generation from natural language descriptions. A plan-and-execute agent workflow first generates execution steps from the natural language description, then iteratively executes each step to construct the diagram. Vision-language models enable direct interpretation of existing diagrams; \citet{koziolek2023llmcodegen} demonstrated GPT-4V generating PLC control code from P\&ID images.

\subsection{HAZOP and Safety Analysis}

Hazard and Operability (HAZOP) studies are critical but time-consuming. \citet{elhosary2024hazop} reviewed HAZOP software tools and proposed AI-assisted frameworks. \citet{lee2026llmhazop} systematically evaluated LLM-based HAZOP: while models achieved high worksheet similarity (F1 $>$ 86\%), only 19--37\% of scenarios were semantically valid. The consensus is that LLMs serve as supportive tools requiring expert oversight \citep{single2020hazop}.

\subsection{Critical Assessment}

Process design applications demonstrate a maturation pattern: early work established representations and feasibility; current work focuses on integration with industrial tools. Key themes include hybrid architectures combining LLM reasoning with structured knowledge (ontologies, knowledge graphs, simulation tools); human-AI collaboration for augmenting rather than replacing engineers; and persistent data limitations requiring proprietary industrial validation. Integration with commercial CAD systems and automated design rule checking represent near-term opportunities.

LLM-enabled design workflows increasingly connect representation, synthesis, and simulation; text-to-simulation agents are discussed in Section 5 as part of process modeling and simulation.
\section{Molecular Design and Synthesis} \label{sec:molecular}

Chemistry has seen impressive LLM demonstrations. \citet{jablonka2024leveraging} showed that fine-tuned GPT-3 can match or exceed specialized ML models for property prediction tasks. Coscientist \citep{boiko2023coscientist} and ChemCrow \citep{bran2024augmenting} combine GPT-4 reasoning with robotic automation to autonomously synthesize compounds. Mobile robotic chemists \citep{burger2020mobile} and end-to-end frameworks \citep{ruan2024llmrdf} extend this paradigm.

Domain-specific models include ChemLLM \citep{zhang2024chemllm} with benchmarks covering nine chemistry tasks, and ChemLLMBench \citep{guo2023chemllmbench} evaluating models on property prediction, yield prediction, and retrosynthesis. General-purpose models like Galactica \citep{taylor2022galactica} and Text+Chem T5 \citep{christofidellis2023textchemt5} show capability across multiple chemistry tasks.

For PSE, these chemistry examples are most relevant because they show how LLMs can orchestrate specialized tools while leaving validation to domain-specific methods. LLMs orchestrate workflows while domain tools verify actions. However, autonomous chemistry operates in favorable conditions (batch timescales, tolerance for failed experiments) that do not always apply to continuous process control.

\section{Process Modeling and Simulation} \label{sec:modeling}

Process modeling and simulation form the computational backbone of PSE, enabling engineers to predict system behavior and optimize operations before physical implementation. \citet{du2025potential} review LLM agent systems for process simulation, proposing frameworks for automated modeling, intelligent design, and bridging theoretical models with industrial application.

\subsection{Natural Language Interfaces}

A barrier to simulation tool adoption is expertise required for configuration. LLM interfaces can lower this barrier: ``Run sensitivity analysis on reactor temperature from 350 to 400°C'' is translated to simulation commands, executed, and results presented accessibly. Several commercial vendors are developing such interfaces for tools like Aspen Plus.

\subsection{Integration with Digital Twins}

\citet{gill2025digital} integrate LLM agents with digital twins for fault management, using high-fidelity simulations to validate proposed actions before physical execution. This validation-before-action paradigm, where LLMs propose and simulations verify, represents a sensible architecture for industrial deployment.

\subsection{End-to-End Process Simulation}

A growing line of work connects LLMs with process simulators, positioning the LLM as an interface or orchestration layer around established simulation tools. \citet{tian2026texttosimulation} developed a multi-agent workflow translating textual process descriptions directly to executable simulations. The Simona dataset provides approximately 1,000 process descriptions of varying complexity for training and evaluation. Agents handle component specification, unit operation selection, parameter inference, and iterative debugging when simulations fail to converge.

\citet{tan2026distillation} demonstrated reasoning agents for distillation process simulation, optimization, and carbon accounting. The agent autonomously collects relevant literature, configures Aspen Plus simulations, optimizes operating conditions, and evaluates decarbonization strategies including heat pump integration. Combining heat pump-assisted processes with renewable energy reduced carbon emissions by 98\% compared to coal-based traditional distillation.

\subsection{Critical Assessment}

LLMs do not provide numerically reliable or formally validated solutions to differential equations; these require purpose-built simulators. The LLM role is orchestration and interpretation: configuring simulations, explaining results, synthesizing across runs. Claims of LLMs ``for simulation'' typically mean LLMs invoking simulation tools, an important distinction.

\section{Time Series Forecasting} \label{sec:forecasting}

Time series forecasting, predicting future values of process variables from historical observations, is fundamental to PSE applications including model predictive control, demand planning, and predictive maintenance. The emergence of transformer-based foundation models for time series represents a significant development that merits dedicated attention.

\subsection{Zero-Shot Forecasting with Language Models}

A surprising result from recent research is that pre-trained language models can perform time series forecasting without fine-tuning on the target dataset. \citet{gruver2024llmtime} demonstrated that by encoding time series as strings of numerical digits, GPT-3 and LLaMA-2 can extrapolate sequences at levels comparable to, and on some benchmarks exceeding, purpose-built forecasting models. The approach leverages LLMs' inherent biases toward repetition and pattern completion. The same properties that enable language completion also capture seasonality and trends in numerical sequences.

The method, called LLMTime, requires no fine-tuning: time series values are converted to text (e.g., ``3.14, 3.15, 3.17, ...''), the LLM generates continuation tokens, and outputs are parsed back to numbers. This zero-shot capability suggests that knowledge learned from text about patterns, periodicity, and sequential dependencies transfers to numerical sequences. LLMs can also naturally handle missing data (represented as special tokens) and incorporate textual side information about the time series context.

\subsection{Reprogramming Approaches}

Rather than treating time series as raw text, reprogramming approaches learn transformations that map time series into representations LLMs can process effectively. \citet{jin2024timellm} introduced Time-LLM, which segments time series into patches and learns to ``reprogram'' these patches into the embedding space of a frozen language model. The LLM backbone (Llama, GPT-2) remains unchanged; only lightweight adapter layers are trained.

Time-LLM achieves state-of-the-art results on standard benchmarks while requiring far less training data than models built from scratch. The approach demonstrates that pre-trained language models contain transferable knowledge about sequential patterns, even though they were never trained on numerical data. Crucially, Time-LLM can incorporate textual prompts describing the dataset characteristics (e.g., ``monthly air passenger data with yearly seasonality''), enabling the model to leverage domain knowledge expressed in natural language.

\subsection{Foundation Models for Time Series}

Inspired by the success of foundation models in NLP, researchers have developed large-scale pre-trained models specifically for time series. These models are trained on massive collections of time series data and can perform zero-shot forecasting on new, unseen datasets. Several comprehensive surveys and comparisons have emerged to characterize this rapidly evolving landscape.

Chronos \citep{ansari2024chronos}, developed by Amazon, tokenizes time series values using scaling and quantization into a fixed vocabulary, then trains T5-family transformer architectures on these tokens using cross-entropy loss. Pre-trained on diverse public datasets augmented with synthetic data from Gaussian processes, Chronos models (ranging from 20M to 710M parameters) demonstrate strong zero-shot generalization: on a 42-dataset benchmark, they achieve comparable and occasionally superior performance to methods trained specifically on each dataset.

TimesFM \citep{das2024timesfm}, developed by Google, takes a decoder-only approach with approximately 200M parameters trained on over 100 billion time points from diverse domains. TimesFM processes time series in patches and generates forecasts autoregressively, similar to how GPT generates text. The model achieves competitive accuracy with statistical methods (ARIMA, ETS) and deep learning baselines while requiring no task-specific training.

Moirai \citep{woo2024moirai}, developed by Salesforce, introduces a masked encoder-based universal forecasting transformer trained on the Large-scale Open Time Series Archive (LOTSA) comprising 27 billion observations across nine domains. Moirai addresses cross-frequency learning through multiple patch size projection layers and handles arbitrary numbers of variates through a novel any-variate attention mechanism. Recent extensions incorporate mixture-of-experts architectures for improved specialization \citep{liu2024moiraimoe}.

Lag-Llama \citep{rasul2024lagllama} focuses on probabilistic forecasting, outputting distributions rather than point predictions. Built on a decoder-only transformer architecture and trained on hundreds of millions of time points, Lag-Llama provides uncertainty quantification essential for risk-sensitive applications in process industries. TimeGPT \citep{garza2024timegpt} represents a proprietary alternative trained on over 100 billion data points. MOMENT \citep{goswami2024moment} provides a family of open time-series foundation models designed for forecasting, classification, anomaly detection, and other downstream time-series tasks.

\subsection{Supervised Transformer Architectures}

Distinct from pre-trained foundation models, supervised transformer architectures require task-specific training but have established strong baselines for time series forecasting. Informer \citep{zhou2021informer} introduced the ProbSparse self-attention mechanism to reduce quadratic complexity, enabling efficient processing of long sequences. Autoformer \citep{wu2021autoformer} incorporates seasonal-trend decomposition directly into the architecture. FEDformer \citep{zhou2022fedformer} applies attention in the frequency domain via Fourier and wavelet transforms.

More recent architectures include PatchTST \citep{nie2023patchtst}, which treats contiguous time points as patches analogous to vision transformers, and iTransformer \citep{liu2024itransformer}, which inverts the standard formulation by treating independent time series as tokens rather than time steps. These supervised approaches typically outperform foundation models when sufficient training data is available for the specific task, but lack the zero-shot generalization capability.



\subsection{Critical Assessment}

For PSE applications, foundation models offer zero-shot capability (no separate training per variable), cross-domain pattern transfer, and multimodal context integration. However, current evaluations use public datasets (energy, traffic, weather) that may not reflect industrial process characteristics. Foundation-model forecasts should be checked against process constraints, since the models do not inherently enforce mass balances, energy balances, or other physical relationships. The pragmatic recommendation is to evaluate these models as baselines rather than replacements, with hybrid physics-constrained approaches meriting investigation for applications requiring interpretability.

\section{Process Optimization and Scheduling} \label{sec:optimization}

Process optimization relies on mathematical programming approaches including linear programming, nonlinear programming, mixed-integer nonlinear programming (MINLP), and dynamic optimization. Classical PSE optimization spans process synthesis, heat and mass integration, supply chain planning, production scheduling, and dynamic operation, providing a broad landscape where LLMs may assist problem formulation and decision support while leaving numerical solution to established frameworks. LLM-based approaches differ fundamentally by operating at the problem formulation and orchestration layer rather than the numerical solution layer.

\citet{zeng2025llmguided} present a multi-agent framework where LLMs autonomously infer operating constraints from minimal process descriptions, then guide optimization. Specialized agents handle constraint generation, parameter validation, simulation execution, and optimization guidance. On a hydrodealkylation case study, the framework achieved competitive solutions with substantially reduced computational effort compared to exhaustive search approaches.

Current evidence suggests that the most promising optimization applications are not direct optimization by LLMs, but rather support functions surrounding optimization workflows: automatic formulation from natural language descriptions, extraction of constraints from technical documentation, generation of algebraic model structures, explanation of results, and orchestration of simulation-optimization loops. In this role, LLMs augment existing solvers rather than replace them. Many industrial problems are formulated as MINLPs whose solution relies on specialized deterministic or decomposition-based algorithms that remain outside the capabilities of current LLMs.

Process scheduling presents unique challenges because feasible solutions must simultaneously satisfy equipment availability, material balances, storage constraints, sequence-dependent changeovers, and production targets. In the broader operations research literature, \citet{cetinkaya2025heuristics} demonstrate that LLMs can discover effective dispatching heuristics for single-machine scheduling benchmarks. Whether such approaches can handle the combinatorial complexity of industrial batch and continuous process scheduling, particularly under uncertainty and frequent rescheduling requirements, remains largely untested.

Overall, current evidence supports a bounded role for LLMs in process optimization. They can assist with formulation, constraint extraction, solver orchestration, and explanation, but dedicated optimization algorithms remain necessary for solving gradient-based, mixed-integer, and dynamic optimization problems. Process scheduling is a particularly important gap: existing demonstrations on simplified scheduling benchmarks do not yet establish capability for industrial batch or continuous operations, where uncertainty, changeovers, storage limits, and frequent rescheduling dominate.
\section{Process Control} \label{sec:control}

``LLM-based control'' has generated significant interest and requires careful examination. Recent surveys provide context for machine learning in industrial control \citep{lawrence2024mlcontrol}.

\subsection{Formal Control-Theoretic Context}

Consider a standard control formulation: given a dynamic system $\dot{x} = f(x, u)$ with state $x$ and input $u$, feedback control seeks a policy $u = \kappa(x)$ that stabilizes the system and satisfies constraints. Classical methods (PID, MPC) admit well-established analysis frameworks: stability margins for PID under modeling assumptions, explicit constraint handling for MPC when the optimization is feasible. LLMs cannot reliably implement $\kappa(\cdot)$ for real-time control; they are too slow, stochastic, and lack formal analysis frameworks. The appropriate role is supervisory: LLMs operate on setpoints, mode selections, or constraint modifications that are then executed by certified control systems.

\subsection{Agentic Control Frameworks}

\citet{vyas2024autonomous} developed an agentic framework for fault recovery and continuous control. Finite State Machines define safe operating envelopes; an LLM proposes recovery sequences; a simulation agent validates transitions; invalid plans trigger reprompting. On laboratory temperature control (TCLab), performance was comparable to PID control.

\citet{guo2024controlagent} present ControlAgent for automating control system design, integrating LLMs with domain expertise to iteratively tune controller parameters for stability, performance, and robustness requirements. The LLMPC framework interprets structured prompting as a form of MPC, arguing that LLMs implicitly minimize planning cost functions \citep{maher2025llmpc}. Related work on vision-language models for MPC \citep{long2024vlmmpc} demonstrates multimodal integration for autonomous systems.

\subsection{Multi-Agent Systems}

Multi-agent LLM frameworks are emerging for complex process operations. \citet{wu2024autogen} developed AutoGen, enabling multi-agent conversation frameworks that can be adapted to process control applications. The wastewater treatment domain has seen particular interest, with multi-agent systems proposed for optimizing operations across biological, chemical, and physical processes. It should be noted that real-world deployment of multi-agent LLM systems in safety-critical process control has not yet been demonstrated at industrial scale.

\subsection{LLM-MPC Hybridization}

Beyond supervisory roles, several hybridization opportunities merit investigation: LLM-assisted model identification from operational logs and experiment descriptions; LLM-generated constraint specifications from safety documentation; LLM-based controller tuning recommendations; and LLM interpretation of MPC solutions for operator communication. These applications leverage LLM strengths (language, knowledge synthesis) while preserving MPC guarantees for real-time execution.

\subsection{Relationship to Reinforcement Learning}

Deep reinforcement learning (RL) has established methods for learning control policies from interaction \citep{badgwell2018rl,spielberg2019selfdriving}. LLM-based agents differ fundamentally in that RL learns $\kappa(\cdot)$ through numerical optimization of cumulative reward; LLMs reason symbolically over action descriptions. Hierarchical RL decomposes control into high-level planning and low-level execution, superficially similar to LLM supervisory architectures, but differs in that both levels are learned numerically rather than combining symbolic reasoning with validated controllers. Safe RL and offline RL address deployment constraints that also apply to LLM agents. The emerging direction of LLM-guided RL, where LLMs provide reward shaping, or policy priors, may combine strengths of both paradigms.

\subsection{Critical Assessment}

Demonstrating ``comparable to PID'' performance is not compelling: PID controllers are simple, fast, well-understood, and backed by decades of tuning theory. If LLMs achieve similar performance with higher latency, complexity, and no stability guarantees, the value proposition is unclear.

\textit{Quantitative comparison}: A tuned PID controller executes a control calculation typically in microseconds to sub-milliseconds on industrial hardware, with deterministic, repeatable behavior and well-characterized stability margins (given appropriate modeling assumptions). Model Predictive Control solves constrained optimization in 10--100 ms, providing explicit constraint satisfaction when the optimization is feasible. LLM inference requires on the order of 500--5000 ms per query\footnote{Order-of-magnitude estimates based on typical cloud API response times for GPT-4 class models; local deployment of smaller models can reduce latency but with reduced capability.}, exhibits stochastic variation across calls, and provides no formal stability or constraint analysis framework. For control loops requiring tens-of-milliseconds response times, LLMs cannot execute fast enough for direct closed-loop control.

\textit{Where LLMs add value}: The genuine opportunity lies in what PID cannot do: interpreting unstructured fault scenarios, coordinating multi-step recovery, integrating diverse information. Using LLMs as supervisory agents can leverage the strengths of both approaches: conventional controllers provide fast, well-characterized low-level control, while LLMs support flexible high-level reasoning over operating modes, procedures, and abnormal situations. In this role, LLMs would propose setpoint changes or mode transitions that are validated and executed by conventional control systems. Because such supervisory decisions typically occur on timescales of minutes to hours, LLM latency is less restrictive than it would be for direct closed-loop control.

\section{Fault Detection and Diagnosis} \label{sec:fdd}

Fault diagnosis may offer LLMs' clearest PSE value, as it is inherently language-rich and existing methods handle interpretation poorly.

\subsection{Time Series Processing}

\citet{qaid2024fdllm} adapted LLMs for machine fault diagnosis from sensor data, demonstrating generalization across operational conditions. SigLLM \citep{alnegheimish2024sigllm} uses LLMs to detect anomalies through forecasting, comparing predictions to actual signals to identify deviations. Similar reprogramming strategies to those used for forecasting \citep{jin2024timellm} could potentially be adapted for fault diagnosis. The LLM-TSFD framework \citep{zhang2025llmtsfd} provides a human-in-the-loop fault diagnosis method for industrial time series using large language models.

\subsection{Multimodal Fault Diagnosis}

\citet{alsaif2024multimodal} propose a multimodal LLM-based fault detection framework featuring a hybrid online/offline architecture and synthetic data augmentation to enhance diagnostic accuracy. Knowledge graph approaches enhance LLM reasoning for fault diagnosis: \citet{liu2024knowledge} embedded aeronautical assembly knowledge graphs into LLMs for online fault diagnosis achieving 98.5\% accuracy. The ability to correlate numerical anomalies with maintenance notes, alarm logs, and operating procedures (information traditionally siloed) is a genuine LLM strength.

\subsection{Explanations and Decision Support}

Beyond detection, LLMs provide natural language explanations: ``Temperature in R-101 exceeded threshold. Similar patterns occurred January 15 during preheater fouling. Recommend checking pressure drop.'' This contextual synthesis, connecting current observations to historical precedents, is difficult with rule-based systems but natural for LLMs. Agentic systems like Argos \citep{gu2025argos} use multiple LLM-powered agents to autonomously generate, verify, and refine interpretable rule-based detectors.

\subsection{Critical Assessment}

LLMs should not replace dedicated fault detection algorithms (PCA, contribution plots) for anomaly detection in structured sensor data. Their value lies in the next stage of the workflow: explaining detections, connecting anomalies to operational context, and suggesting possible follow-up actions. Architectures that retrieve specific maintenance records, alarm logs, or operating procedures alongside the LLM interpretation allow operators to verify the basis for each recommendation. Because these recommendations may influence safety-relevant decisions, they should remain advisory and subject to human and procedural oversight \citep{amodei2016concrete}.

\subsection{Illustrative Scenario: Reactor Temperature Excursion}

Consider a concrete workflow illustrating the architecture of Figure~\ref{fig:architecture}. During night shift, reactor R-101 experiences an unexpected temperature excursion. The operator queries the system: ``What caused the temperature spike in R-101?''

The LLM orchestration layer: (1) retrieves recent temperature, pressure, and flow trends from the historian; (2) queries the alarm log for related events; (3) searches maintenance records and incident reports for similar patterns; (4) identifies that a comparable excursion occurred three months prior during catalyst regeneration. The system responds: ``Temperature increased 15°C over 20 minutes starting at 02:47. Cooling water flow dropped 12\% at 02:45, similar to the March 15 incident caused by partially closed valve CV-103. Recommend checking CV-103 position and cooling water header pressure.''

The operator requests validation: ``What if we increase cooling flow by 20\%?'' The system invokes a digital twin simulation, confirms the proposed action would restore temperature within 8 minutes without violating constraints, and presents the result. The operator approves; the conventional control system executes the setpoint change with full safety interlocks active.

This workflow illustrates the division of responsibility that is likely to be most practical in industrial settings: the LLM supports diagnosis and proposes possible actions, simulation checks feasibility, certified control systems execute approved changes with existing safeguards, and the human operator remains accountable for the final decision.

\section{Cross-Cutting Challenges} \label{sec:challenges}

\subsection{Hallucination}

LLMs generate confident falsehoods, a critical concern for safety applications. In PSE contexts, hallucinations may include fabricated reaction mechanisms, incorrect safety limits, or non-existent equipment tag references. Unlike grammatical errors, such fabrications can be difficult for non-experts to detect and potentially dangerous if acted upon. Mitigation strategies (RAG, output validation, human oversight) help but add complexity. Applications must assume occasional errors and include verification appropriate to consequences.

\subsection{Latency and Deployment}

Frontier models require seconds per query via cloud APIs. Smaller local models offer lower latency but reduced capability. Papers demonstrating results with GPT-4 while ignoring deployment constraints are incomplete. For context: a well-tuned PID controller typically executes in microseconds to sub-milliseconds; MPC solves optimization problems in 10--100 milliseconds; LLM inference requires on the order of 500--5000 milliseconds. This difference of several orders of magnitude fundamentally constrains real-time applications.

\subsection{Safety Certification Barriers}

Process industries operate under rigorous functional safety standards. Safety Instrumented Systems (SIS) must achieve specified Safety Integrity Levels (SIL), requiring demonstrated failure rates, deterministic behavior, and extensive validation evidence. LLMs present fundamental certification challenges: probabilistic outputs that vary with temperature settings and random seeds; opaque decision processes that resist systematic failure mode analysis; model updates that may silently change safety-relevant behavior; and cloud deployment models that introduce network dependencies and version control complexity. 

At present, no established pathway exists to certify an LLM as part of a safety function at SIL 1 or higher. This does not preclude LLM deployment in advisory roles with human oversight, but it does constrain autonomous safety-critical applications. Research on formal verification of neural networks and guaranteed uncertainty quantification may eventually provide certification pathways, but these remain distant prospects.

\subsection{Industrial Data Realities}

Academic demonstrations typically assume clean, well-labeled datasets. Industrial process data presents challenges that can undermine LLM performance: missing signals from sensor failures or maintenance; calibration drift introducing systematic errors; inconsistent tag naming across plant areas or equipment vintages; IT/OT network separation restricting data access; proprietary formats from legacy DCS and historian systems; and sparse labeling of fault events or operational modes.

The data distribution in published papers differs substantially from industrial operations. Most demonstrations use public benchmarks (Tennessee Eastman, CSTR simulations) that, while valuable, cannot capture the heterogeneity of real plant data. Transfer learning and domain adaptation techniques show promise but require further validation on proprietary industrial datasets that academic researchers rarely access.

\subsection{Validation Frameworks}

Process industries have mature frameworks for control system validation \citep{qin2003survey}. Nothing comparable exists for LLM-based systems. How to test, characterize failure modes, and ensure model updates do not change safety-relevant behavior are open questions requiring community attention. The classic reviews of fault detection methods \citep{venkatasubramanian2003review,venkatasubramanian2003reviewII,venkatasubramanian2003reviewIII} provide relevant context for understanding what LLM approaches must ultimately match or exceed.

\subsection{Evaluation Practices and Benchmarks}

Evaluation practices remain uneven across LLM-related PSE studies. Retrosynthesis and time-series forecasting have more established benchmark cultures, while HAZOP, control, and industrial fault diagnosis often rely on proprietary cases, simulations, or laboratory demonstrations. This fragmentation makes cross-domain comparison difficult and highlights the need for representative process-industry benchmarks. Few papers provide industrial validation; most rely on public benchmarks or laboratory demonstrations. Dataset availability is mixed: some benchmarks are public, but industrial datasets remain proprietary. The community would benefit from standardized evaluation protocols and representative industrial benchmarks.

Synthesizing across the approximately 70 works reviewed: the majority ($>$80\%) rely on public benchmarks or simulation environments; fewer than 10\% report validation on proprietary industrial data; laboratory demonstrations (e.g., TCLab, robotic chemistry platforms) account for a small fraction of experimental work. No papers report sustained deployment in production process control systems. Dominant metrics vary by domain (top-$k$ accuracy for synthesis tasks, MSE/MAE for forecasting, F1 for fault detection) but cross-domain comparison remains difficult due to inconsistent reporting practices.

\subsection{Data Governance and Security}

Industrial deployment introduces data governance challenges beyond technical performance. Cloud-based LLM APIs raise data exfiltration concerns: process data, operating procedures, and incident reports may contain proprietary information inappropriate for external transmission. On-premise deployment addresses this but requires substantial infrastructure. Access control and auditability are essential: who can query what data, and how are LLM recommendations logged for post-incident analysis? Prompt injection attacks pose risks in RAG and agentic setups: malicious content in retrieved documents could manipulate LLM behavior. Model versioning and change control matter for safety-critical contexts: updates that silently change recommendations require regression testing protocols analogous to those for control system software. Human factors also merit attention: operator trust calibration, cognitive load from additional information streams, alarm fatigue from frequent recommendations, and clear accountability when LLM-assisted decisions lead to incidents.

\subsection{Integration Architecture}

Practical deployment requires integration with existing DCS, SCADA, and MES systems. A viable architecture positions LLMs in advisory and orchestration roles: (1) conventional controllers executing validated control laws; (2) a data layer aggregating historian, alarm, and event data; (3) an LLM orchestration layer generating recommendations; and (4) a human interface layer where operators retain decision authority.

\subsection{Reproducibility}

The reproducibility of LLM-based research is complicated by model versioning: results obtained with GPT-4-0314 may not replicate with GPT-4-turbo, yet few papers specify exact model versions or API timestamps. Prompt sensitivity compounds this challenge; minor wording changes can substantially alter outputs. The community would benefit from standardized reporting requirements including model identifiers, temperature settings, and representative prompt examples.
\section{Future Directions} \label{sec:future}
Table~\ref{tab:maturity} synthesizes the maturity landscape across PSE domains. 
Time series forecasting and molecular design have reached benchmark maturity with standardized evaluation protocols. Process design, simulation, and fault diagnosis show pilot-stage demonstrations. Process control and optimization remain early-stage, reflecting the difficulty of integrating probabilistic language model outputs with domains requiring precise numerical guarantees.

\begin{table}[t]
\centering
\caption{Maturity landscape of LLM applications across PSE domains.}
\label{tab:maturity}
\footnotesize
\setlength{\tabcolsep}{3pt}
\begin{tabular}{@{}p{2.1cm}ccp{2.4cm}@{}}
\toprule
\textbf{Domain} & \textbf{LLM Role} & \textbf{Maturity} & \textbf{Evidence} \\
\midrule
Process design & \cellcolor{MedColor!60}Med & \cellcolor{PilotColor!60}Pilot & Flowsheets, P\&IDs, HAZOP \\[2pt]
Molecular design & \cellcolor{HighColor!60}High & \cellcolor{BenchColor!60}Bench/Pilot & ChemBench, retrosyn. \\[2pt]
Simulation & \cellcolor{MedColor!60}Med & \cellcolor{PilotColor!60}Pilot & Text-to-sim agents \\[2pt]
Time series & \cellcolor{HighColor!60}High & \cellcolor{BenchColor!60}Bench & Chronos, TimesFM, Moirai \\[2pt]
Optimization & \cellcolor{MedColor!60}Med & \cellcolor{EarlyColor!60}Early & Multi-agent, sched. heuristics \\[2pt]
Process control & \cellcolor{LowColor!60}Low & \cellcolor{EarlyColor!60}Early & TCLab, supervisory agents \\[2pt]
Fault diagnosis & \cellcolor{HighColor!60}High & \cellcolor{PilotColor!60}Pilot & FD-LLM, multimodal FDD \\
\bottomrule
\end{tabular}
\vspace{4pt}
\parbox{\columnwidth}{\footnotesize\textit{LLM Role}: High = core method, Med = significant component, Low = supervisory or advisory role only. \textit{Maturity}: Early = conceptual/lab, Bench = standardized evaluation, Pilot = industrial PoC.}
\end{table}

Several directions merit sustained research investment. Natural language interfaces for process operations address genuine industrial needs: operators could query historians, retrieve procedures, or explore what-if scenarios through conversation rather than navigating complex software interfaces. Domain-specific models like ChemLLM \citep{zhang2024chemllm} demonstrate that chemical knowledge can be embedded effectively, though whether PSE applications require purpose-built models or can leverage general LLMs with retrieval augmentation remains an open question \citep{schweidtmann2024generative,woo2025llmpse}.

Agentic architectures represent perhaps the most promising near-term direction. The pattern of coupling LLM reasoning with simulation-based validation \citep{liang2026llmagentchemsim} addresses the reliability concerns that have limited LLM deployment in safety-critical settings. Multi-agent frameworks \citep{wu2024autogen} offer natural decomposition for complex workflows spanning multiple tools and domains. The key insight is that LLMs need not be trusted to produce correct answers directly; instead, they can propose actions that are verified by established numerical methods before execution.

\subsection{Research Themes}

Five research themes merit sustained investigation. First, \textit{trustworthy agent architectures} that combine LLM reasoning with formal verification of proposed actions could enable deployment in higher-stakes settings. Second, \textit{physics-grounded reasoning} through neuro-symbolic approaches or physics-informed prompting may reduce hallucinations involving conservation laws and experiment constraints. Third, \textit{industrial benchmark datasets} representative of real plant conditions, with appropriate anonymization, would accelerate progress beyond public benchmarks. Fourth, \textit{human-AI collaboration} research should address trust calibration, cognitive load, and effective division of responsibility between operators and AI advisors. Fifth, \textit{certification and validation frameworks} adapted for advisory AI systems would provide regulatory pathways currently absent.

Specific research questions include: Can retrieval-augmented LLMs achieve sufficient reliability for unattended advisory roles? What minimum local model size preserves useful capability for on-premise deployment? How should operator trust be calibrated when LLM confidence does not correlate with accuracy?

\subsection{Industrial Copilots}

The nearest-term commercial applications may be industrial copilots: conversational interfaces for querying historians, retrieving procedures, assisting shift handovers, and supporting maintenance decisions. These applications leverage LLM strengths in natural language while operating in advisory roles where errors have limited immediate consequences. Several vendors have announced or deployed such systems, though published evaluations remain sparse.

\subsection{Timeline Assessment}

Based on current trajectories, natural language interfaces and documentation assistants appear deployable within 1--2 years; agentic workflows with digital twin validation within 3--5 years; semi-autonomous supervision, if achievable, likely requires a decade or more of reliability advances and regulatory evolution.

\subsection{Workforce Implications}

LLM adoption will reshape PSE education and workforce expectations. Engineers may increasingly serve as system integrators rather than algorithm developers, requiring curricula balancing traditional mathematical foundations with AI literacy. Conversely, over-reliance on LLM-generated solutions risks eroding deep domain expertise that remains essential for novel situations.
\section{Conclusions}

This survey has reviewed LLM applications across PSE, organizing literature into process design, molecular synthesis, modeling, time series forecasting, optimization, control, and fault diagnosis. We find genuine promise in applications leveraging LLM strengths (natural language processing, knowledge synthesis, flexible reasoning) while applications requiring real-time performance or formal guarantees remain challenging.

The most impactful near-term work is architectural: designing systems where LLMs augment rather than replace human judgment and traditional methods. The conceptual contribution of this survey is that LLMs are best positioned as reasoning and integration layers above validated numerical methods, serving as orchestrators and interfaces rather than as substitutes for physics-based tools. This requires collaboration between AI researchers understanding LLM capabilities and process engineers understanding industrial requirements. As foundation model capabilities continue to advance, periodic reassessment of this landscape will be warranted; we anticipate that the architectural principles identified here, LLMs as orchestrators rather than replacements, will remain valid even as specific capabilities evolve. We hope this survey, and the accompanying IFAC World Congress workshop/tutorial, contribute to that collaboration.

\section*{Declaration of Generative AI and AI-Assisted Technologies in the Writing Process}

During the preparation of this article the authors used Claude (Anthropic) to assist with drafting and editing text. After using this tool, the authors reviewed and edited the content as needed and take full responsibility for the content of the publication.

\small
\bibliography{references_v1}
\end{document}